\documentclass[twocolumn]{aastex631}

\begin{document}

\title{The Evolution of Substructure during Star Cluster Assembly}

\correspondingauthor{Alison Sills}
\email{asills@mcmaster.ca}

\author[0009-0008-6388-1630]{Edwin Laverde-Villarreal}
\affiliation{Departamento de Física, Universidad Nacional de Colombia, Carrera 45 No. 26-85, Edificio Uriel Gutiérrez, Bogotá D.C., Colombia}
\affiliation{Department of Physics and Astronomy, McMaster University, 1280 Main Street West, Hamilton, ON, L8S 4M1, Canada}

\author[0000-0003-3551-5090]{Alison Sills}
\affiliation{Department of Physics and Astronomy, McMaster University, 1280 Main Street West, Hamilton, ON, L8S 4M1, Canada}

\author[0000-0002-6116-1014]{Claude Cournoyer-Cloutier}
\affiliation{Department of Physics and Astronomy, McMaster University, 1280 Main Street West, Hamilton, ON, L8S 4M1, Canada}

\author{Veronica Arias Callejas}\affiliation{Departamento de Física, Universidad Nacional de Colombia, Carrera 45 No. 26-85, Edificio Uriel Gutiérrez, Bogotá D.C., Colombia}

\begin{abstract}

Star cluster formation and assembly occurs inside filamentary and turbulent molecular clouds, which imprints both spatial and kinematic substructure on the young cluster. In this paper, we quantify the amount and evolution of this substructure in simulations of star cluster formation that include radiation magnetohydrodynamical evolution of the gas, coupled with detailed stellar dynamics, binary formation and evolution, and stellar feedback. We find that both spatial and kinematic substructure are present at early times. Both are erased as the cluster assembles through the formation of new stars as well as the merger of sub-clusters. Spatial substructure is erased over a timescale of approximately 2.5 times the initial free-fall time of the cloud. Kinematic substructure persists for longer, and is still present to the end of our simulations. We also explored our simulations for evidence of early dynamical mass segregation, and conclude that the presence of a population of binary stars can accelerate and enhance the mass segregation process. 
\end{abstract}


\section{Introduction}

Star formation is an inherently clustered process, in which molecular clouds simultaneously form many stars as they collapse \citep{LadaLada}. The turbulent nature of these clouds produces a complex filamentary structure in the molecular gas. Those filaments themselves collapse and fragment, but also funnel material to hubs where filaments come together \citep[e.g.][]{Kirk13, Wells24}. The stars inherit the structure and kinematics from the gas out of which they formed. Embedded star-forming regions exhibit a wide range of morphologies ranging from chains and clumpy substructure to spherical relaxed systems \citep{Kuhn2014}. At some point in the star formation process, the molecular cloud has formed enough stars that stellar dynamics becomes another important physical process that drives the evolution of the stellar system. Ultimately, the combination of star formation and feedback will remove the gas from the system. At that point, some fraction of the newly-formed stars will disperse into the field while others may remain bound as a stellar cluster. The conditions under which bound clusters are formed, and the dependence of the properties of those clusters on their birth environment, remains an open question. 

The key to answering these questions is the relative timescales of the important physical processes. From the perspective of simulations, this should be considered to be somewhat equivalent to the choice of initial conditions. There is a long history of stellar dynamics simulations which aim to predict the sizes, densities, boundedness, and stellar population properties such as mass segregation of stellar clusters after formation and gas expulsion. Some of these simulations assume that the cluster is monolithic, spherical, and bound as the gas is removed from the system \citep[e.g.][]{Proszkow2009,FariasTan23} while others implicitly assume a hierarchical cluster assembly process by choosing fractal initial conditions \citep[e.g.][]{Goodwin2004} or using initial conditions from hydrodynamics simulations of molecular cloud collapse \citep[e.g.][]{Ballone2021}. While all these simulations agree that molecular cloud properties such as surface density and initial virial parameter are important, the detailed structure, kinematic properties, and stellar distributions depend strongly on what initial configurations are chosen. 

A consensus is emerging that at early times, clusters assemble hierarchically. 
Large-scale simulations of dense molecular clouds \citep[e.g.][]{Howard2018,Grudic22} suggest that massive clusters are formed through the merger of many smaller sub-clusters. Smaller-scale simulations of molecular cloud collapse and star formation with an accurate treatment of stellar and binary dynamics also see continuous mergers of subclusters, but also find evidence for complex dynamical evolution that can include splitting of subclusters \citep{Cournoyer-Cloutier2023}. 
Stellar dynamics models which started with the observed properties of present-day young cluster-forming regions showed that, even in the presence of the remaining gas, highly substructured regions would quickly evolve into spherical configurations \citep{Sills2018}. Therefore, a simulation which started at a slightly later time might appropriately use a monolithic initial condition. However, the differences between the stellar dynamics simulations suggest that it is important to understand cluster formation and assembly in detail over the first few tens of millions of years. 

On the observational side, we are now able to probe the spatial and kinematic structure of young clusters and cluster complexes in the Milky Way in exquisite detail. A blind search of the Gaia DR3 data release finds thousands of clusters \citep{HuntReffert2023}, with accompanying information such as sizes, distances, ages, stellar membership, and stellar velocities from proper motions. For those objects which were also part of spectroscopic surveys such as Gaia-ESO, we have radial velocities as well \citep{Wright2024}. These and similar datasets have been used to show that star formation events can span tens of parsecs and can take the form of cluster chains \citep{Posch2025}, that some open clusters show evidence of spatial subgroups or substructure \citep{GregorioHetem2024}, and that young open clusters can show both spatial and kinematic substructure which diminishes with inferred dynamical age \citep{Arnold24}. Therefore, we can start to use these observations to address the question of timescales, and to constrain the types of dynamical models which best describe the formation and early assembly of star clusters.

Another tracer of dynamical evolution of star clusters is mass segregation i.e. the tendency for massive stars to be more centrally concentrated than lower-mass stars. Mass segregation can be primordial, if massive stars are formed near the centre of the cluster \citep[e.g.][]{Bonnell06}. Mass segregation is also an outcome of two-body relaxation \citep{BinneyTremaine} which occurs on a cluster's dynamical (i.e. two-body relaxation) timescale. This timescale is typically longer than the age of young embedded clusters, so observational evidence of mass segregation is usually interpreted as support for primordial segregation. However, small sub-clusters have much shorter dynamical times than larger clusters, and so it is possible that dynamical mass segregation can be present at young ages, and it can also be dynamically enhanced when these sub-clusters merge to form a larger main cluster \citep{McMillan07}. More recent simulations of hierarchical star cluster formation including gas dynamics \citep{Polak25} also link increased dynamical mass segregation to the overall dynamical environment and timescales over which that environment changes. 

In this paper, we use simulations of star cluster formation which include both detailed radiation hydrodynamics, including stellar feedback processes, and accurate star-by-star stellar dynamics. We quantify the spatial and kinematic substructure in star-forming regions and follow their evolution as the cloud continues to form stars, as the subclusters evolve, and as stellar feedback ejects the natal gas. We use the same statistical tests which have been applied to observed embedded star clusters so that we can compare to the recent Gaia results for local clusters. By having the full simulation data of positions, velocities, and evolutionary histories, we can quantify the evolutionary histories of these star clusters and compare to the observational results. This will allow us to quantify the importance of the physical processes we see in our simulations, and to confirm the validity of the observational inferences about hierarchical vs monolithic star cluster assembly. 

In section 2 we describe the simulations and the statistical tests we use, as well as our cluster selection process in the data from the simulations. Section 3 gives the results for our simulated cluster-forming regions, which are compared to observational results and the implications are discussed in section 4. 

\section{Methods}

\subsection{Star Cluster Formation Simulations}
We use the stellar masses, positions, and velocities from the star cluster formation simulations presented in~\citet{Cournoyer-Cloutier2024}. The initial conditions and star formation metrics of the simulations are outlined in Table~\ref{tab:simulations}. Those simulations use the cluster formation code \textsc{Torch}~\citep{Wall2019, Wall2020}, in which the \textsc{Amuse} framework~\citep{PortegiesZwart2009, Pelupessy2013, PortegiesZwart2013, PortegiesZwart2019} couples magnetohydrodynamics to stellar dynamics, star and binary formation, and stellar mechanical and radiative feedback. 

\begin{table}[tb!]
    \begin{center}
    \begin{tabular}{ccccc}
    GMC & $M_{\mathrm{gas}}$ [M$_{\odot}$] & $\Sigma$ [M$_{\odot}$ pc$^{-2}$] & $t_{\mathrm{ff}}$ [Myr] & $M_{*}$ [M$_{\odot}$] \\
    \hline
    M1 & 2 x 10$^4$ & 130 & 1.06 & 7.8 x 10$^3$\\
    M2 & 8 x 10$^4$ & 520 & 0.530 & 3.44 x 10$^4$ \\
    M3 & 3.2 x 10$^5$ & 2080 & 0.265 & 1.95 x 10$^5$\\
    \hline
    \end{tabular}
    \end{center}
    \caption{Initial conditions, and stellar mass formed in the simulations. Columns: Cloud label, initial gas mass, initial surface density, initial free-fall time, and mass in stars formed. All clouds have an initial radius of 7 pc.}
    \label{tab:simulations}
\end{table}

\begin{figure*}[tb!]
    \includegraphics[width=\linewidth, clip=true, trim=0.5cm 3.5cm 0cm 2.5cm]{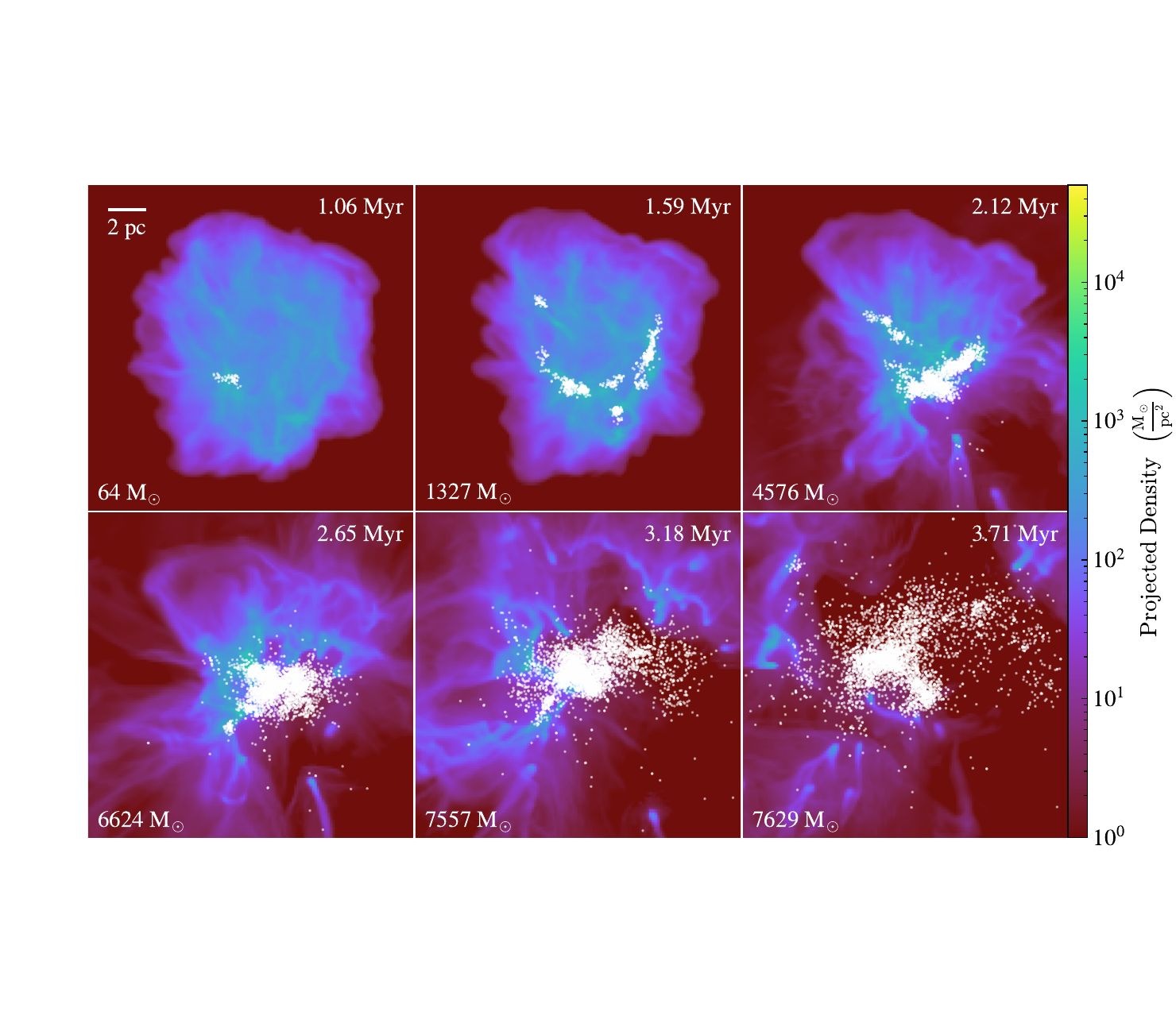}
    \caption{Gas column density in colour, with the stars shown as white markers, for simulation M1 after 1.0, 1.5, 2.0, 2.5, 3.0 and 3.5 free-fall times. The physical time of the simulation is given in the top right corner, and the total mass in stars is given in the bottom left corner.}
    \label{fig:gas_evolution}
\end{figure*}

Magnetohydrodynamics, star formation, radiation, and mechanical feedback are handled by \textsc{Flash}~\citep{Fryxell2000, Dubey2014}, and we adopt a resolution of 0.137 pc at the highest refinement level of our adaptive grid. Star formation in our simulations is handled by sink particles, which form from bound, dense gas in regions where the gas flows are locally converging, following the criteria outlined in~\citet{Federrath2010}. Upon formation, each sink particle samples a~\citet{Kroupa2002} initial mass function from 0.40 to 150 M$_{\odot}$ and a mass-dependent binary sampling algorithm~\citep{Cournoyer-Cloutier2024} based on observed binaries~\citep{Moe2017, Offner2023} to generate a list of systems to be formed. After is has formed, a sink particle continuously accretes gas from its surroundings. As it accretes enough mass to form the next system on the list, the system is spawned from the sink: its mass is removed from the sink particle and given to star particles, conserving mass locally. The system's center of mass is placed randomly within the sink accretion radius, with a velocity drawn from a normal distribution centered on the sink velocity with a standard deviation corresponding to the local sound speed. This process is repeated as the sink particle accretes more gas throughout the simulation. The details of binary formation from sink particles are outlined in~\citet{Cournoyer-Cloutier2021}. We handle stellar dynamics, including hard binaries and close encounters, with the N-body code \textsc{PeTar}~\citep[][see \citealt{Polak2024} for the implementation in \textsc{Torch}]{Wang2020b}. Stars more massive than 13 M$_{\odot}$ influence the surrounding gas through momentum-driven winds, ionizing radiation, and direct radiation pressure~\citep{Wall2020}. The code also contains prescriptions for supernova feedback, but none of our simulations evolve for long enough to have any massive stars explode. The time evolution of the gas and stars in our lowest-mass cloud is shown in Fig.~\ref{fig:gas_evolution}.

\subsection{Cluster identification and membership criteria}

\begin{figure*}[tb!]
    \centering
    \begin{minipage}{.49\textwidth}
        \centering
        \includegraphics[width=\linewidth, clip=True, trim=2.5cm 1cm 1cm 2cm]{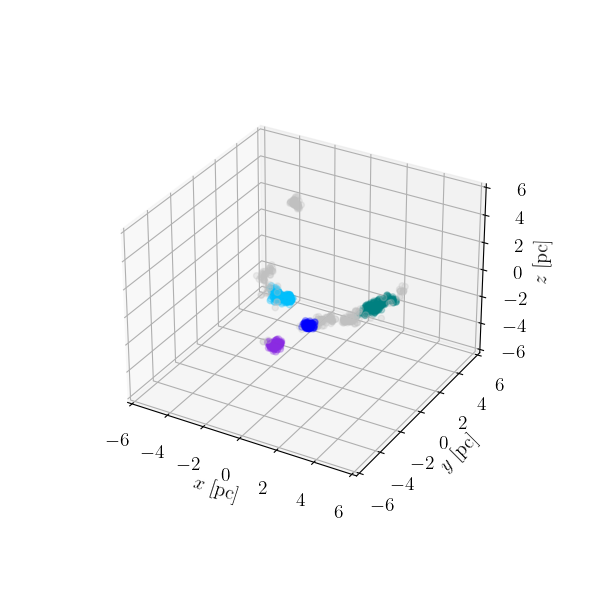}
    \end{minipage}%
    \begin{minipage}{0.49\textwidth}
        \centering
        \includegraphics[width=\linewidth, clip=True, trim=2.5cm 1cm 1cm 2cm]{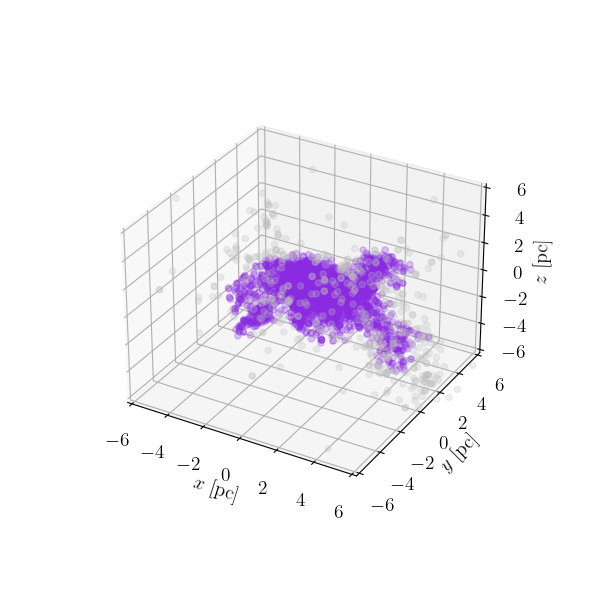}
    \end{minipage}
    \caption{Clusters identified in M1 after 1.5 (left) and 3.0 (right) initial free-fall times of the cloud. Each cluster is shown as one colour, and unclustered stars are shown in grey. Examples were chosen to showcase a time at which there is a lot of substructure and one at which only one cluster was identified. Clusters can have any morphology, and clearly show evidence of spatial asymmetries that persist at late times.}
    \label{fig:dbscan}
\end{figure*}

We identify clusters with DBSCAN~\citep{Ester1996, Pedregosa2011} based on the stars' three-dimensional positions. We require a star to have at least five neighbors~\citep[optimal for three-dimensional data, ][]{Sander1998} within a distance $d$, where the distance is calculated from the knee in the distribution of distances to the fifth nearest neighbor for each snapshot using \texttt{kneed}~\citep{Satopaa2011}. An example of the clusters identified at two times in M1 is shown in Fig.~\ref{fig:dbscan}. At early times, star formation is happening only in the densest gas and so we see a very substructured region with small clusters well-separated from each other. At later times, the ongoing star formation as well as the motions of the stars away from their birth location has created a single, but clearly not spherical, star cluster.

\begin{figure}[htb!]
    \includegraphics[width=\linewidth]{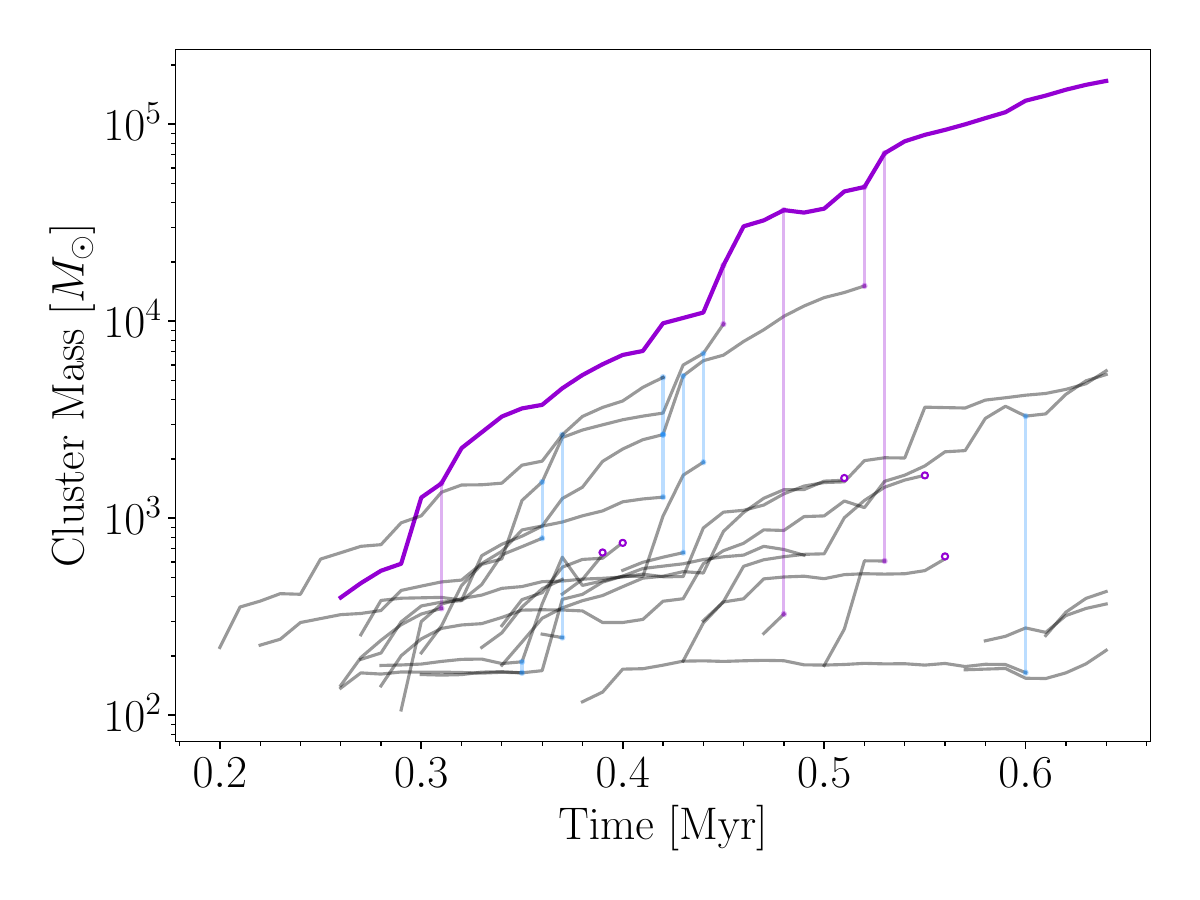}
    \caption{Total cluster mass as a function of time for each of the clusters identified in simulation M3. The most massive cluster at the latest snapshot is shown in violet. The vertical blue lines show instances where a smaller cluster merged with a larger cluster, with the violet ones representing mergers with the most massive cluster. The evolution lines with an open marker at their end show clusters that eventually ended up becoming part of the most massive cluster but did so through a complicated process of splitting and merging with other clusters, so their details are not shown. The final most massive cluster was not the first cluster formed, nor was it the most massive when it formed, but through a combination of continued star formation and mergers of smaller clusters, it comes to encompass most of the stellar mass in the simulation.
    \label{fig: merger_tree}}
\end{figure}

We can follow the evolution of the clusters throughout the simulations, following the method described in~\citet{Cournoyer-Cloutier2023}. For each cluster, we track its main progenitor, identified as the cluster in the previous snapshot with which it shares the largest fraction of its mass. We also follow mergers between clusters, which can happen when 2 or more previously single clusters get close enough together that DBSCAN detects them as being part of a single larger cluster. This enables us to study the time evolution of the clusters formed in each simulation. Clusters can grow in mass through ongoing star formation, and by merging with smaller clusters. They can lose mass through expansion, star ejection due to dynamical events, or by stars just being loosely bound to the cluster and easily leaving it. The evolution and merging histories of these clusters can be seen better through a merger tree plot. Since we can uniquely identify each cluster that appears throughout the simulation and follow its evolution, we can plot the number of stars that each cluster has at each snapshot. Additionally, using the merging information, we can join certain evolution lines at the times at which the respective clusters merge to better appreciate the dynamics of the simulation. An example of such a merger tree for our most massive simulation, M3, is shown in Figure~\ref{fig: merger_tree}.

In this simulation, clusters of initially a few hundred stars are formed in regions where the gas is densest. Some of the smaller clusters merge into larger clusters, shown as vertical lines in the figure. The violet line represents the most massive cluster at the end of the simulation, which is assembled through a combination of ongoing star formation, mergers with smaller clusters, cluster splitting, and mergers with clusters that themselves are the products of mergers. This behavior is consistent with our understanding of star formation in locally very dense regions in the cloud at early times, particularly along dense filaments, and was described in detail in \citet{Cournoyer-Cloutier2023}. As the cloud evolves and the gas is removed both by stellar feedback and by ongoing star formation, the stars fill more of the available volume and eventually come to form one main cluster. At the end of this simulation, the most massive cluster contains a little more than half the total stellar mass, with the next largest clusters being more than a factor of 10 less massive. The evolution of our smaller clouds, M1 and M2, is similar but because there are fewer stars and less star formation, the merger trees are simpler and show less hierarchical merging structure. The complexity of the assembly process for clusters clearly results in spatial substructure over some of the cluster formation process, and as we will show, also results in kinematic substructure which persists for longer. 

Our simulations include a significant population of binary stars, which must be treated carefully during our analysis of substructure. The metrics of spatial and spatial-kinematic substructure described in the following sections rely on the positions and velocities of individual cluster members. For the kinematic statistics in particular, the method we employ relies on comparisons between the direction of motion of a star and its closest neighbors. In the case of binary stars, however, the direction of motion of a star and that of its companion will be exactly opposite as they orbit their shared center of mass. To use those statistical measurements despite the high binary fraction of our simulated clusters, we identify binaries with semi-major axes smaller than 10,000 au~\citep[as done in][]{Cournoyer-Cloutier2024} and replace the two binary stars with a single star having the total mass of the binary, its center of mass position and center of mass velocity for our subsequent analysis. This approach not only removes artificial over-estimates of kinematic or spatial substructure caused by binary systems, but is more consistent with observations which usually cannot resolve the individual components of a binary system. 

We also follow the approach of 
\citet{Arnold24} to remove velocity outliers from the analysis. Before determining the spatial-kinematic substructure for 48 clusters, they remove all stars which have a speed that is more than 2.5 times the interquartile range away from the median speed (equivalent to excluding stars more than approximately 3$\sigma$ from the centre of a Gaussian distribution). When applying these criteria to our simulations, we remove fewer than 1\% of the stars in our clusters under investigation. We also tested more relaxed criteria and confirm that only the most extreme outliers (beyond 25 times the interquartile range away from the median speed) affect our kinematic substructure statistic. The formation of runaway stars due to binary interactions can cause a sharp change to the kinematic substructure statistic which does not describe the overall evolution of the cluster. 

\subsection{Spatial Substructure}

A commonly-used parameter to measure the spatial substructure (or the lack thereof) of a distribution of points is the Q statistic \citep{Q04}. This dimensionless quantity is calculated as the ratio between two quantities related to the spatial distribution of the points: $Q = \bar{m} / \bar{s}$. The quantity $\bar{m}$ is the normalized mean edge length of the minimum spanning tree (MST) of a 2D projection of the distribution. In our simulations, we calculated the Q parameter using each of the xy, xz, and yz projections and report the mean of all three. There is only a small spread between the three projections, as expected since our initial simulations have no preferred geometry. The MST is the collection of edges that join the points in the distribution, ensuring that every point is reachable from every other point in a unique way (i.e. no loops), while also minimizing the total sum of all edges. The value $\bar{s}$ is the normalized mean separation distance between the points. In our analysis, we normalize the mean separation of all the stars by dividing it by the normalization parameter $(N-1)/\sqrt{N \pi}$ \citep{Q04}, where $N$ is the total number  of stars in the cluster. In this way, the Q parameter can range from values of 0 up to arbitrarily large numbers, although most reported values in real clusters tend to be less than 1. Values close to 0 would mean a small $\bar{m}$ compared to $\bar{s}$, which corresponds to a highly subclustered distribution, where the points arrange in tight isolated groups (small $\bar{m}$) that are separated from each other (large $\bar{s}$). Larger values of Q would come from a smooth, uniform distribution, where $\bar{m}$ is similar to $\bar{s}$. In general, we define Q values below 0.8 to represent substructured distributions, and those above it to represent smooth centrally-concentrated distributions.

\subsection{Kinematic Substructure}

It has been shown that kinematic substructure can be studied by means of Moran's I statistic \citep{Arnold22}, and that this statistic is reliable under certain observational biases on the data (high uncertainties on stellar velocities, contamination of the star sample, and low completeness). Moran's I describes the degree of autocorrelation of a variable $\mu$, in this case a velocity. It essentially indicates if similar $\mu$ values appear near each other in space (positive I, correlation), if dissimilar $\mu$ values are near each other (negative I, anti-correlation), or if there is no structure regarding the arrangement of $\mu$ (I is close to zero). The threshold for Moran's I values that corresponds to no substructure is not exactly zero, but rather $\frac{-1}{N-1}$, where N is the number of stars (approximately $10^4$ to $2\times10^5$ for our simulations). 

For calculating Moran's I, we need to have the $\mu$ value for each of the $N$ points in space, and also to compute a weight $w_{ij}$ for each pair of points $i \neq j$. These weights are taken as the inverse of the physical distance between the corresponding points, and are then normalized by rows, i.e. $w_{ij} = \frac{w'_{ij}}{\sum_j w'_{ij}}$, where $w'_{ij}$ are the un-normalized weights. With this, Moran's I is calculated as follows:
    \begin{equation}
        I(\mu) = \frac{N}{\sum_{i \neq j} w_{ij}} \frac{\sum_{i \neq j} w_{ij} (\mu_i - \overline{\mu})(\mu_j - \overline{\mu})}{\sum_{i} (\mu_i - \overline{\mu})^2}
        \label{eq: MoransI}
    \end{equation}

To analyze the kinematic substructure in the detected clusters over the evolution of our simulations, we focus on the Moran's I values for the velocities, specifically $I(v_{x})$, $I(v_{y})$, and $I(v_{z})$. Moreover, following \citet{Arnold22}, we can average two of these values to obtain a less noisy metric that reduces the influence from the arbitrary choice of frame of reference. This combined value, $I(v_{\text{2D}})$, encapsulates the information from both 2D velocities, providing a clearer overall measure of the cluster's kinematic substructure. Just as in our spatial substructure analysis, we calculate $I(v_{\text{2D}})$ for three cases to take into account the effect of choice of frame of reference. We calculate $I(v_{\text{2D}})$ for the three combinations of pairs of velocities, namely ($v_x$,$v_y$), ($v_x$,$v_z$), and ($v_y$,$v_z$). Again, there is only a small spread between the three values of $I(v_{\text{2D}})$ due to the lack of a preferred axis in our initial conditions. It is worth noting that we use the 3D positions to determine the distances between the points for the weights.

\subsection{Mass Segregation}

The most widely accepted way to measure the mass segregation present in a star cluster is through the mass segregation ratio $\Lambda_{\text{MSR}}$, which was first proposed by \citet{Allison09}. This statistic compares the MSTs of two distributions: that of the $N_{\text{MST}}$ most massive stars, and that of $N_{\text{MST}}$ randomly selected stars. We first have to determine the total edge length of the massive group's MST, $l_{\text{mass}}$. We then need to randomly select a high number of sets of $N_{\text{MST}}$ non-massive stars (we will use 500 sets), and calculate the average total length $\left< l_{\text{norm}} \right>$ of their MSTs. The standard deviation of the distribution of these random MSTs can be interpreted as the dispersion of the average length, $\left< l_{\text{norm}} \right> \pm \sigma_{\text{norm}}$. Once we have these values, we can calculate $\Lambda_{\text{MSR}}$ as the ratio between the two lengths:

\begin{equation}
    \centering
    \Lambda_{\text{MSR}} = \frac{\left< l_{norm} \right>}{l_{mass}} \pm \frac{\sigma_{norm}}{l_{mass}}
\end{equation}

In this way, values of $\Lambda_{\text{MSR}}$ which are greater than 1 represent the presence of mass segregation, while those less than 1 represent inverse mass segregation (massive stars are less centrally concentrated than lighter ones), and values close to 1 denote no signal of mass segregation.

Despite its common use, $\Lambda_{\text{MSR}}$ was developed primarily for small-N star clusters, so we will turn most of our attention to a recently-proposed modification of this statistic by \citet{Wei25}. The main change in this new statistic has to do with the value of $l_{\text{mass}}$. To determine this value, we randomly sample a fixed number of stars (in our case, 25) from the total number of massive stars, and do this many times. In this paper, we select 3000 subgroups. For each subgroup, we calculate the total edge length of its MST. The average of these MST lengths across all subgroups gives us the mean edge length for the massive stars, denoted by $\left< l_{\text{mass}} \right>$. With this new value, the modified $\Lambda_{\text{MSR}}$ is then calculated as follows:

\begin{equation}
    \Lambda_{\text{MSR}} = \frac{\left< l_{\text{norm}} \right>}{\left< l_{\text{mass}} \right>} \pm \frac{\langle l_{\text{norm}} \rangle}{\langle l_{\text{mass}} \rangle} \sqrt{  \left( \frac{\sigma_{\text{norm}}}{\langle l_{\text{norm}} \rangle} \right)^2  + \left( \frac{\sigma_{\text{mass}}}{\langle l_{\text{mass}} \rangle} \right)^2 }
\end{equation}

The interpretation for the values of this statistic remains the same as for the original version. Using this modified version we get rid of the dependence with $N_{\text{MST}}$, and can report a single value for our clusters at each snapshot (as opposed to doing so for a range of $N_{\text{MST}}$ values). This version of the statistic is more robust for populous clusters, as it can sample over a large number of stars while still retaining the original understanding developed for small-N systems. In the original paper, \citet{Wei25} apply this version of the statistic to the massive (a few times $10^4$ M$_{\odot}$), young (less than 10 Myr) Milky Way cluster  Westerlund 1. They obtain a value of $\Lambda_{\text{MSR}}$ of 1.11 $\pm$ 0.11, suggesting a small amount of mass segregation in that cluster. 

\section{Results}

For our simulations, we quantify both the spatial and kinematic substructure over time of the forming star clusters. In this section, we concentrate on the most massive clusters, as those are the objects which have the most complex assembly histories and are most likely to show any evidence for, and evolution of, substructure. For each simulation, we perform two analyses: first, we calculate the Q and I statistics for the most massive cluster present in each snapshot. Secondly, we identify those stars which make up the most massive cluster in the final snapshot, and then calculate their Q and I statistics in each of the previous snapshots. The first analysis is equivalent to observing clusters in the Milky Way, while the second analysis directly follows the assembly and evolution of what will become the dominant cluster produced by the giant molecular cloud.

We also analyze the mass segregation present along the evolution of the final biggest cluster. Additionally, we assess the effect that the presence of primordial binaries has in this process. For this, we analyze two groups of stars, one where we take all the stars as they are, the raw group, and one where we reduce the binary systems to their center-of-mass properties, the reduced group. Lastly, we compare our results with those obtained by other researchers using simulations constructed under the same frameworks, but without a primordial binary prescription.

\subsection{Spatial Substructure}

\begin{figure*}[t!]
    \epsscale{1.15}
    \plotone{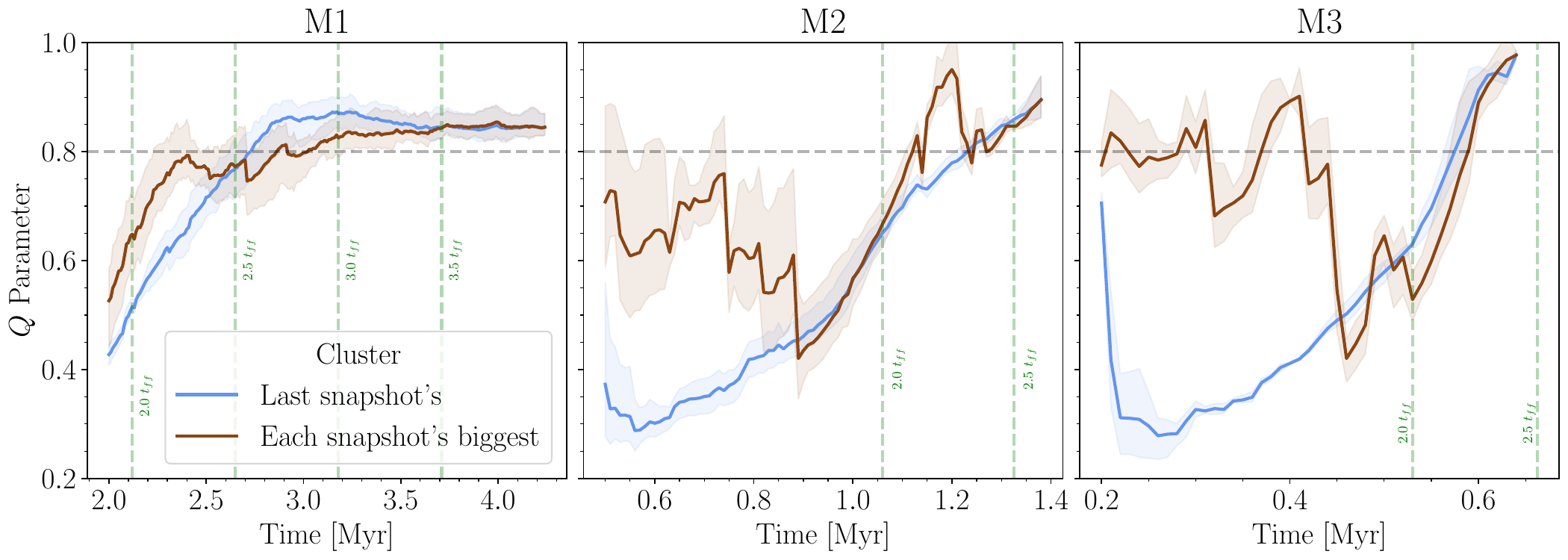}
    \caption{Q parameter in time for 2 cases. The brown line is found using the stars in the most massive cluster at each snapshot, and the blue line using the stars that are part of the most massive cluster in the last snapshot. The Q value of 0.8 helps distinguish between a substructured spatial distribution ($<$0.8), and a smooth centrally concentrated one ($>$0.8). The green lines denote multiples of that cloud's free fall time, starting at 2.0 for the leftmost line, and increasing in 0.5. Note that the evolution of M2 and M3 only extends to about 2.5 free fall times. Spatial substructure has been lost in both M1 and M2, at times that roughly coincide with 2.5 times the cloud's initial free fall time.
    \label{fig: Q_comparison}}
    \end{figure*}

We first analyze the spatial substructure of clusters in our simulations. The Q values are plotted along the evolution of the clusters present in the 3 simulations in Figure \ref{fig: Q_comparison}. The 0.8 threshold that separates smooth and substructured distributions is shown. The brown line shows the Q value for each snapshot's most massive cluster, and the blue line shows the stars present in the most massive cluster in the latest snapshot. We see an erratic behavior in the brown line, including sudden increases and decreases, going above and below the 0.8 threshold. This behavior is expected as we are not always looking at the same cluster from snapshot to snapshot, and the properties of the cluster will change dramatically at a moment when it mergers with another cluster. The blue line shows a smoother evolution, as expected since we are following the same stars throughout the simulation, from the moment they form until the end of the simulation. In all cases, as soon as the most massive cluster is more than a few hundred solar masses, the amount of both spatial substructure generally diminishes with time.  We note that the time at which the Q values cross the 0.8 threshold, i.e. the time at which the spatial substructure signature is lost, roughly coincides with 2.5 free-fall times for all clusters.

    \subsection{Kinematic Substructure}

\begin{figure*}[ht!]
    \epsscale{1.15}
    \plotone{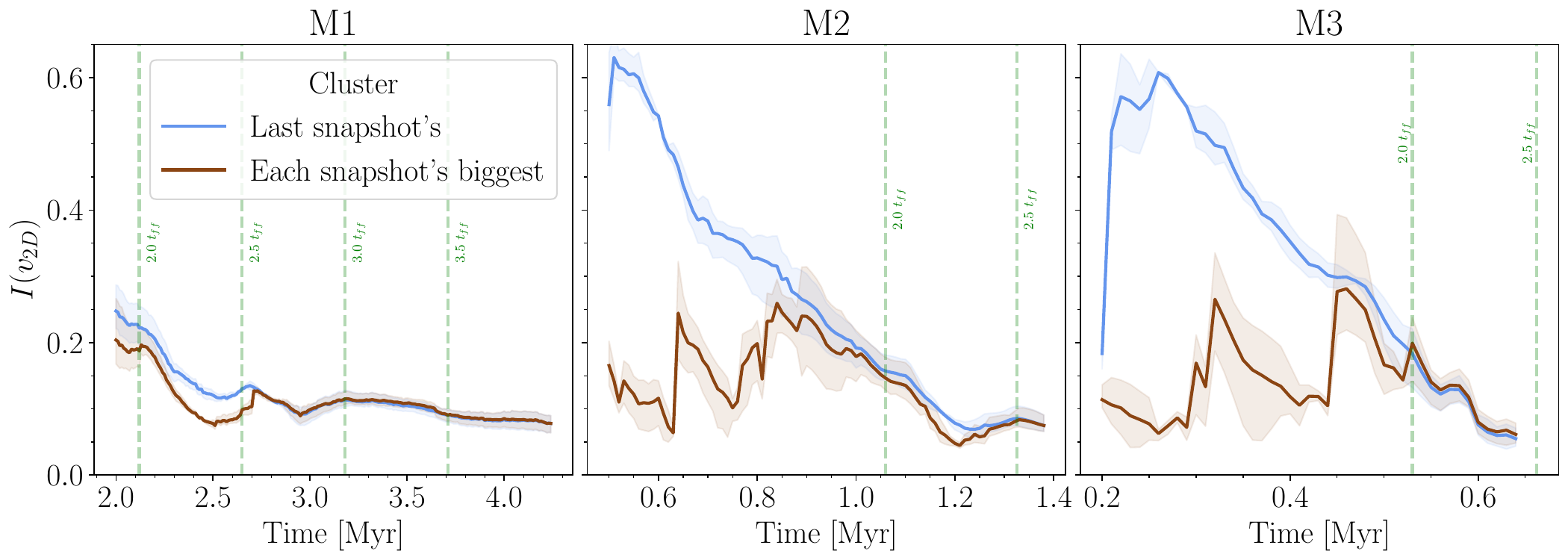}
    \caption{Moran's I values in time for the three simulations. The brown line was calculated using the most massive cluster at each snapshot, and the blue line using the stars that belong to the most massive cluster in the latest snapshot. The green lines denote multiples of that cloud's free fall time, starting at 2.0 for the leftmost line, and increasing in 0.5. At early times the behavior is rather erratic due to the constant merging of clusters; as time passes, the evolution is more steady, namely I decreases steadily, which indicates the cluster is losing kinematic substructure as it evolves.
    \label{fig: MI_both}}
    \end{figure*}

We now look at the kinematic substructure. In Figure \ref{fig: MI_both} we plot Moran's I values against time for the 3 simulations. Just as for Q, the brown line is found using each snapshot's most massive cluster, and the blue line shows the evolution of I for the stars from the most massive cluster in the latest snapshot. The presence of significant merger events are obvious as the sharp increases in I. When a group of new stars becomes part of the most massive cluster, they bring with them the bulk velocity of their cluster which is often different than that of the cluster, but that difference is erased over time as the two subpopulations mix dynamically. In addition, mergers of subclusters result in a small fraction of the stars to become unbound \citep{Karam2024} which will also serve to reduce the kinematic substructure of the cluster. The evolution of the stars which will eventually become the most massive cluster (blue line) is smoother, although small indications of subcluster mergers can be seen. As in the case of the spatial substructure, the system evolves towards an unstructured configuration, but unlike the spatial substructure, the systems do not reach that point (I=0) during our simulation time. In fact, the systems seem to be asymptoting towards an I value of 0.1, suggesting that there should be persistent, but low-level, kinematic substructure in young clusters before dynamical relaxation has had a chance to erase such structure. The larger clouds have initially more kinematic substructure compared to the lowest mass simulation M1, but after between 2.5 and 3 freefall times, all three simulations have similar values of kinematic substructure.

\subsection{Mass Segregation}

\begin{figure*}
\epsscale{1.15}
\plotone{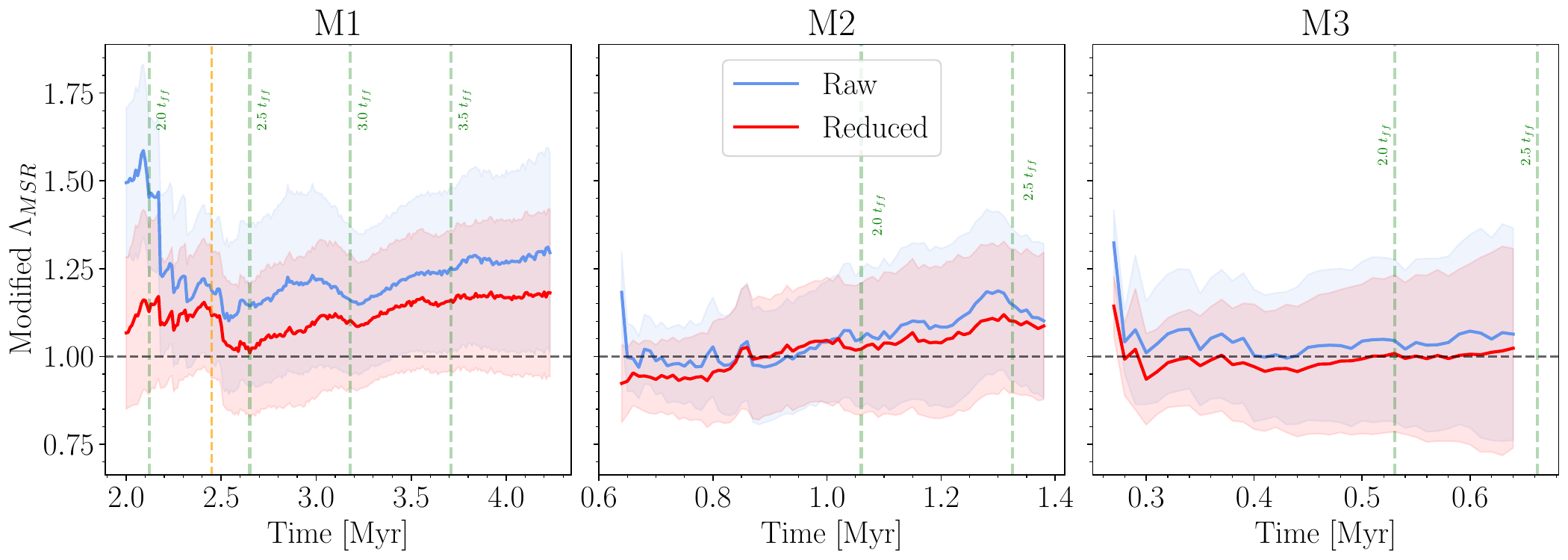}
\caption{Modified mass segregation ratio as a function of time for all three simulations. The shaded regions indicate the dispersion of the statistic. The dashed green lines denote multiples of that cloud's initial free fall time, starting at 2.0 for the leftmost line, and increasing in 0.5. The dashed orange line marks the time of collapse of M1. The blue lines marked `raw' use the actual positions and masses of all stars, whether they are in binary systems or not; the red `reduced' lines replace each binary system with an object at the binary's center of mass, with a mass equal to the total mass of the system.}
\label{fig: modMSR}
\end{figure*}

Lastly, we turn our attention to the process of mass segregation. We note that any signature of mass segregation seen in our simulations must be dynamical, as we do not impose any constraints on the formation location of massive stars relative to the eventual centre of the final cluster. In Figure \ref{fig: modMSR} we plot the modified $\Lambda_{\text{MSR}}$ values for the 3 simulations. For this analysis, we treat our binary stars in two different ways. The `raw' analysis treats each component of a binary system as its own object with its own mass and position; this is what could be done in observations of nearby clusters where binaries can be separated. The `reduced' analysis replaces each binary system with a single object at the centre of mass of the system, and with a mass equal to the total mass of the system. Binaries are not able to be resolved in distant clusters.

The values shown are those obtained by considering the stars in the last snapshot's biggest cluster. We can see that there is a relatively high signal of mass segregation across most of the shown evolution of M1. We see a similar behaviour, though with lower signal, for M2 and M3. We can also see that the values for the raw group are generally bigger than those for the reduced one. This makes sense if we consider that most massive stars are part of a binary system \citep{Moe2017}, and since we are replacing these with one object, the total edge length of the MST will inevitably increase as the short edge that joined the binaries will be replaced by a larger one. 

More interestingly, if we focus on M1, we can see that mass segregation is larger before the time of collapse of the cluster (when its stars reach their densest configuration) noted by the dashed orange line. We see a similar, though less strong, behaviour in M2, with the highest values of the mass segregation ratio being followed by a drop happening right before the expected time of collapse (a few snapshots after the end of the simulation). The evolution time for M3 is not enough to see the collapse of the cluster, so we do not see this behaviour. This increase in mass segregation near the collapse of the cluster is consistent with what \citet{Allison09} suggested, namely that dynamical mass segregation can occur during the collapse of a cluster when a dense core with a small crossing time is formed.

If we compare our results with those obtained by \citet{Polak25}, whose simulations were run under the same framework but without the primordial binary prescription, we note two main differences. First, we see that our simulations have a generally stronger signal of mass segregation across the evolution of our clusters. Second, while they also observe a quick increase in mass segregation, it happens after the time of collapse, whereas we see it before cluster collapse. This then suggests that the inclusion of primordial binaries enhances and accelerates the process of mass segregation. This could be explained by considering that binaries, are more dynamically active than single stars \citep{Heggie75}. Since the gravitational cross section scales as the semi-major axis of the orbit of the binary instead of the radius of the single star, binaries are more likely to interact with other systems, hardening themselves or other binary systems, ejecting low-mass stars from the centre of the cluster, and causing more energy exchange between the stars and the cluster as a whole. 

    \section{Summary and Discussion}

We have investigated the evolution of spatial and kinematic substructure of stars during the process of forming and assembling star clusters. We applied the Q parameter and Moran's I statistic to simulations of star formation in giant molecular clouds that included stellar feedback and allowed for the formation of a population of primordial binaries. All our simulations show spatial and kinematic substructure at early times, and in general the substructure diminishes over a few initial free-fall times of the cloud. The spatial substructure is erased first, with the kinematic substructure remaining, albeit at a low level, throughout our simulations. The behavior of our three simulations is similar, suggesting that the processes which erase early substructure are not dependent on the masses of the initial clouds nor on the number of stars in the clusters.

Lastly, we used a modified version of the $\Lambda_{\text{MSR}}$ statistic to measure the signals of mass segregation throughout the evolution of our simulations. We found a strong signal across the evolution of our lowest-mass simulation, and a less strong one for the other two simulations. In all cases we see that the mass segregation signal increases as the cluster evolves, both before and after the collapse of the cluster, as expected from dynamically-driven mass segregation. We further note that the mass segregation signals in our simulations are stronger and happen earlier than in simulations without a primordial binary prescription. Thus, we conclude that the presence of primordial binaries both enhances and accelerates the process of mass segregation.

We note that our simulations still have some residual gas left over from the star formation process, and in particular simulations M2 and M3 have not yet finished forming all the stars that will make up the final clusters. We also note that the cluster assembly process is likely not complete. The clusters are, at most, a few Myr old. Their observational counterparts investigated by, for example, \citet{Wright2024} and \citet{Arnold24} range in age from a few to a few tens of Myr. Future work should include the dynamical evolution of these systems after gas expulsion, for example building on the work of \citet{Sills2018} or \citet{McMillan07}, to fully compare these simulations to the observed clusters. However, we are already able to show that the timescales for the different kinds of substructure (spatial, kinematic, and mass segregation) are different. Knowing and characterizing the various kinds of substructure in young star clusters and in regions which are currently assembling star clusters can be an important tool to probe and characterize the hierarchical assembly of star clusters. 

\begin{acknowledgments}
ELV was supported by a Mitacs Globalink Research Fellowship. CCC is supported by a Canada Graduate Scholarship -- Doctoral (CGS D) from the Natural Sciences and Engineering Research Council of Canada (NSERC). AS is supported by NSERC. This research was enabled in part by support provided by Compute Ontario (\url{https://www.computeontario.ca/}) and the Digital Research Alliance of Canada (\url{alliancecan.ca}) via the research allocation FT \#2665: The Formation of Star Clusters in a Galactic Context.

\end{acknowledgments}

\bibliography{Bibliography}{}
\bibliographystyle{aasjournal}

\end{document}